%% file: n-trace.tex
\documentclass[journal=cmatex,manuscript=article,layout=twocolumn]{achemso}

\usepackage[version=3]{mhchem} 
\usepackage{xcolor} 

\input{config.tex}

\newcommand{\revision}[1]{#1}  

\newcommand{\rrevision}[1]{#1}

\author{Chen Zhang}
\email{zhangc@ornl.gov}
\affiliation{Computer Science and Mathematics Division, Oak Ridge National Laboratory, Oak Ridge, TN 37831, USA}

\author{Valerie A. Niemann}
\affiliation{Department of Chemical Engineering, Stanford University, Stanford, CA 94305, USA}
\alsoaffiliation{SUNCAT Center for Interface Science and Catalysis, SLAC National Accelerator Laboratory, 2575 Sand Hill Road, Menlo Park, CA 94025, USA.}

\author{Peter Benedek}
\affiliation{Department of Chemical Engineering, Stanford University, Stanford, CA 94305, USA}
\alsoaffiliation{SUNCAT Center for Interface Science and Catalysis, SLAC National Accelerator Laboratory, 2575 Sand Hill Road, Menlo Park, CA 94025, USA.}

\author{Thomas F. Jaramillo}
\affiliation{Department of Chemical Engineering, Stanford University, Stanford, CA 94305, USA}
\alsoaffiliation{SUNCAT Center for Interface Science and Catalysis, SLAC National Accelerator Laboratory, 2575 Sand Hill Road, Menlo Park, CA 94025, USA.}

\author{Mathieu Doucet}
\affiliation{Neutron Scattering Division, Oak Ridge National Laboratory, Oak Ridge, TN 37831, USA}

\title[nTrace]
{Extracting thin film structures of energy materials using transformers}
\abbreviations{NR, ML}
\keywords{
    Reflectometry,
    Machine Learning,
    Transformer encoder,
    Energy materials,
    Thin films,
}

\begin{document}
\let\thefootnote\relax\footnotetext{%
This manuscript has been authored by UT-Battelle, LLC, under contract DE-AC05-00OR22725 with the US Department of Energy (DOE).
The US government retains and the publisher, by accepting the article for publication, acknowledges that the US government retains a nonexclusive, paid-up, irrevocable, worldwide license to publish or reproduce the published form of this manuscript, or allow others to do so, for US government purposes.
DOE will provide public access to these results of federally sponsored research in accordance with the DOE Public Access Plan (https://www.energy.gov/doe-public-access-plan).
}

\begin{abstract}
Neutron-Transformer Reflectometry Advanced Computation Engine (\ModelName{}), a neural network model using transformer architecture, is introduced for neutron reflectometry data analysis.
It offers fast, accurate initial parameter estimations and efficient refinements, improving efficiency and precision for real-time data analysis of lithium-mediated nitrogen reduction for electrochemical ammonia synthesis, with relevance to other chemical transformations and batteries.
Despite limitations in generalizing across systems, it shows promises for the use of transformers as the basis for models that could \revision{accelerate traditional} approaches to modeling reflectometry data.
\end{abstract}

\section{Introduction}
Neutron reflectometry (NR) has proved to be an important tool to study energy-related materials.
With its isotopic and light element sensitivity, it enables us to investigate mechanisms involving lithium and hydrogen.
The low neutron absorption cross-section for most materials also allows us to build electrochemical cells to study processes in situ, which is crucial to understanding what often are multi-step processes with more than a single time scale.

One challenge practitioners of this technique face is the difficulty of modeling the data in a timely manner.
Solving the inverse problem and extracting structural information from reflectometry data usually requires prior knowledge of the system and how it is expected to behave.
\revision{Such prior knowledge is often gained from complementary characterization techniques.}
For energy-related materials, solid-electrolyte interphase (SEI) type layers can have a wide array of substructures and diffuseness. It can extend from \SI{5}{\nano \meter} to \SI{500}{\nano \meter}.
These surface features are therefore often difficult to predict, which \revision{can make the process} of extracting the thin film structure from electrochemical systems challenging.
These challenges especially apply to novice users, who require substantial training before becoming independent in their NR data analysis.
Modeling requires knowledge of scattering principles to develop the intuition needed to develop physically meaningful models and assess them.

Machine learning (ML) has proved to be a promising tool to help facilitate the process of modeling reflectometry data.
From simple neural networks~\cite{Aoki:2021, Mironov:2021, Doucet_2021}, to variational auto-encoders~\cite{Andrejevic-2022} and reinforcement learning~\cite{Doucet:2024}, extracting structural parameters of thin films using ML was proved to be possible, yet still very dependent on prior knowledge of the system and its properties.
The recent creation of large language models (LLM) like ChatGPT~\cite{openai_gpt4_blog} has brought on the exciting question of whether foundation models for science will soon become reality.
At the heart of these LLMs lies an architecture known as the transformer, pioneered by Vaswani and coworkers~\cite{vaswani_attention_2023}.
This innovation marked a significant advancement in the realm of sequence learning.
Before the advent of the transformer, Recurrent Neural Networks (RNN) were primarily used for Natural Language Processing (NLP) tasks, owing to their adeptness at handling both short and long-term memory~\cite{sherstinsky_fundamentals_2020}.
However, the scalability of RNNs was constrained by their need for prior input to finalize a prediction.
In contrast, transformers employ a multi-head attention mechanism capable of extracting both short-range and long-range patterns from a sequence, simultaneously.
This distinctive trait facilitates the scalability of transformer-based models.

The ability of transformers to discern long-range patterns within input data empowers them to manage complex dependencies and contexts.
This attribute is particularly beneficial for processing natural language and has proved to be equally transformative in other domains.
For instance, beyond its initial success in NLP, the transformer architecture has demonstrated remarkable adaptability, extending its capabilities to fields such as computer vision~\cite{peebles_scalable_2023}, graph reasoning~\cite{fatemi_talk_2023}, audio processing~\cite{verma_audio_2021}, and many more.
This cross-domain versatility highlights the potential of foundational models, similar to LLMs, that could be harnessed as a multi-purpose tool across various scientific disciplines.
Such models could democratize advanced domain knowledge, streamline the scientific discovery process, and propose innovative strategies to address complex challenges that currently cannot be resolved in real time.

\begin{figure}[htp]
    \centering
    \begin{tikzpicture}[node distance=0.45cm and 0.2cm]
          \node[startstop] (start) {Raw measurements};
          \node[process, below=of start] (reduce) {R curve};
          \node[process, below left=of reduce] (conventional) {human};
          \node[process, below right=of reduce] (proposed) {N-TRACE};
          \node[process, below=of conventional] (offline) {R fitter};
          \node[decision, below=of offline] (fitgood) {fit good?};
          \node[process, below=of fitgood] (refiner) {refiner};
          \node[startstop, below=of refiner] (results) {results};
          \node[process, below=of proposed] (inference) {guess};
        
          \draw[arrow] (start) -- (reduce) node [midway, right, label] {reflectivity calculation};
          \draw (reduce) -| (conventional) node [midway, label, xshift=-1.2cm, yshift=-0.1cm] {conventional};
          \draw (reduce) -| (proposed) node [midway, label, xshift=0.82cm, yshift=-0.2cm] {proposed};
          \draw [arrow] (conventional) -- (offline) node [midway, left, label] {guess};
          \draw [arrow] (offline) -- (fitgood)node [midway, left, label] {fit};
          \draw [arrow] (fitgood) -- node[near start, right] {yes} (refiner);
          \draw [arrow] (refiner) -- (results) node [midway, left, label] {evaluate};
          \draw [arrow] (proposed) -- (inference) node [midway, left, label] {inference};
          \draw [arrow] (inference) |- (refiner);
          
          \draw[arrow] (fitgood.west) -- ++(-0.5,0) |- (conventional.west) node [midway, label, yshift=0.2cm] {no};
    \end{tikzpicture}
    \caption{%
    \revision{
    The workflow chart depicts two paths: the conventional offline iterative parameter extraction (left), and the proposed \ModelName{} based on auto extraction.
    The proposed method replaces the labor-intensive iterative search process with a neural network, thus transferring the search work to the training process.
    }
    }
    \label{fig:workflow chart}
\end{figure}

The present work represents a stride in this direction, leveraging the unique attention mechanism of the transformer and its associated long-range pattern recognition capacity to establish a correlation between experimental measurements of thin films and their underlying material structure.
Specifically, we have developed \ModelName{} (Neutron-Transformer Reflectometry Advanced Computation Engine), a neural network model calibrated to predict thin film structure for lithium-mediated nitrogen reduction for ammonia production.\footnote{
\revision{The example implementation of \ModelName{} can be found in this \href{https://code.ornl.gov/hygnn/capabilities/reflectometry/-/blob/main/src/tgreft/models/refl_gpt.py}{repository}.}
}
This system has been the focus of our research for several years and therefore constitutes a use case for which both prior knowledge of the chemistry involved and knowledge of the associated scattering theory is available~\cite{blair-2022,blair-2023}. 
It therefore represents a unique opportunity to develop a physics-informed model using both synthetic and real data.

We envisage a workflow where data is automatically analyzed, helping users steer their experiments and expedite publication (\Cref{fig:workflow chart}).
In our experience, the work presented here would have saved months of data analysis.
The emphasis of this study is the steady-state of the system.
Although there is a significant interest in the dynamics of nitrogen reduction as the SEI is forming with time-resolved measurements, the application of transformers to dynamic data will be the subject of future work.

\section{Experimental}
The data used here were measured as part of an ongoing project to study electrochemical production of ammonia~\cite{niemann-2024}.
The study focuses on understanding the mechanisms involved at the surface of a copper electrode in contact with a non-aqueous electrolyte comprised of deuterated tetrahydrofuran (THF), lithium tetrafluoroborate (LiBF$_4$) salt, and ethanol under constant current.
The details of similar measurements performed with molybdenum instead of copper can be found in this group's previous work~\cite{blair-2022,blair-2023}. 
The data used here were measured using 16 samples in various steady-state conditions.
For each sample, an initial open-circuit voltage (OCV) measurement was performed, followed by OCV measurements that took place after applying a constant current in the range of -2 and -0.1~mA/cm$^2$ for 2 to 5 minutes.
The samples were prepared using physical vapor deposition (PVD).
A titanium layer of approximately 50 \AA{} was deposited on a single crystal silicon substrate, and a layer of approximately 500 \AA{} was deposited on top of the titanium.

Neutron reflectometry measurements were performed at the Liquids Reflectometer~\cite{ankner-2008} (LR) at the Spallation Neutron Source at Oak Ridge National Laboratory. 
The LR is a time-of-flight reflectometer with a wavelength band of about 3.5 \AA{}.
Reflectivity measurements covered a wavevector transfer
$Q=4\pi\sin\theta/\lambda$
in the range of \QSIR{0.009}{Q}{0.18},
where $\theta$ is the scattering angle and $\lambda$ is the wavelength of the neutron.
In total, 51 reflectivity measurements were used for this work.

The NR measurements were modeled using the refl1d~\cite{brian_benjamin_maranville_2020_4329338} package.
For each measurement, the sample structure was modeled using a stack of layers, each with a thickness, a 
scattering length density (SLD) value, and a roughness parameter representing the interface between 
adjacent layers. Together, the stack of layers describes an SLD profile as a function of depth into 
the film, which represent the structure of the film and carries information about its composition.
The SLD can be written as
\begin{equation}
SLD = \sum_i n_i b_i
\end{equation}
which runs over all types of atoms in the system, where $n_i$ is the number density of
element~$i$ and $b_i$ its coherent scattering length. For the measurements performed here, the initial
state can be modeled by adding a copper oxide on top of the copper layer. As current is applied
to the film, two layers rapidly form on the surface: a thinner layer containing lithium species, and
a thicker, diffuse, SEI containing electrolyte degradation products. The transformer-based
model presented here is able to predict structure varying between these two extremes.

\section{Methodology}
This section describes the process for developing \ModelName{}.
This neural network model is designed to analyze and interpret structural parameters from thin film reflectivity measurements.
\ModelName{} consists of a transformer encoder and a multilayer perceptron (MLP) decoder.
The transformer's multi-headed attention system helps to identify patterns from the reflectivity curves.
The MLP then converts these patterns into associated structural parameters.
Key aspects of this method include a unique data preparation and preprocessing approach, a hybrid loss function suited to the model's dual objectives, and a streamlined procedure for training, validation, and evaluation.
Synthetic data are used for training and validation, while actual experimental measurements are used for final evaluation.
\ModelName{} is implemented in PyTorch~\cite{NEURIPS2019_9015} and trained with an Nvidia Tesla V100S-PCIE-32GB card.
The training, validation, and evaluation processes are coordinated using \revision{\href{https://github.com/mlflow/mlflow}{mlflow}}, which enhances the reproducibility and efficiency of research workflows~\cite{chen_mlflow_2020}.

\subsection{Data preparation and preprocessing}

\ModelName{} processes raw reflectivity measurements and interpolates them to a pre-determined $q$ range, specifically from \revision{0.009 1/\AA{} to 0.18 1/\AA{}},
corresponding to the $q$ range of the measurements.
The interpolated reflectivity curve is then weighted according to their $q$ value, 
followed by a log filter (\Cref{fig:enter-label}). 
This ensures the output embedding has adequate variation and a stable value range, both of which are critical for constructing a stable neural network~\cite{Doucet_2021}.
\footnote{%
In the upcoming version of N-TRACE, the interpolation step will be replaced with a graph neural network-based embedding module.
This will automatically convert input reflectivity measurements into a stable embedding within the latent space.
}

\begin{figure}
    \centering
        \begin{tikzpicture}[node distance=2cm, auto]
        \tikzstyle{process} = [rectangle, minimum width=3cm, minimum height=1cm, text centered, draw=black, fill=orange!30]
        \tikzstyle{arrow} = [thick,->,>=stealth]
    
        \node (step1) [process] {$R(q)$};
        \node (step2) [process, below of=step1] {$R'(q), q \in (0.009, 0.18)$};
        \node (step3) [process, below of=step2] {$\Phi$};
    
        \draw [arrow] (step1) -- (step2) node [midway, right, label] {interpolate};
        \draw [arrow] (step2) -- (step3) node [midway, right, label] {$*q_i^2/q_0^2, \log_{10}$};
        \end{tikzpicture}
    \caption{%
    Workflow diagram illustrating \ModelName{}'s data preprocessing steps, converting raw measurements $R(q)$ into latent embedding $\Phi$.
    }
    \label{fig:enter-label}
\end{figure}

\subsection{Model architecture}

\begin{figure}[htp]
\centering
\includegraphics[width=.45\textwidth]{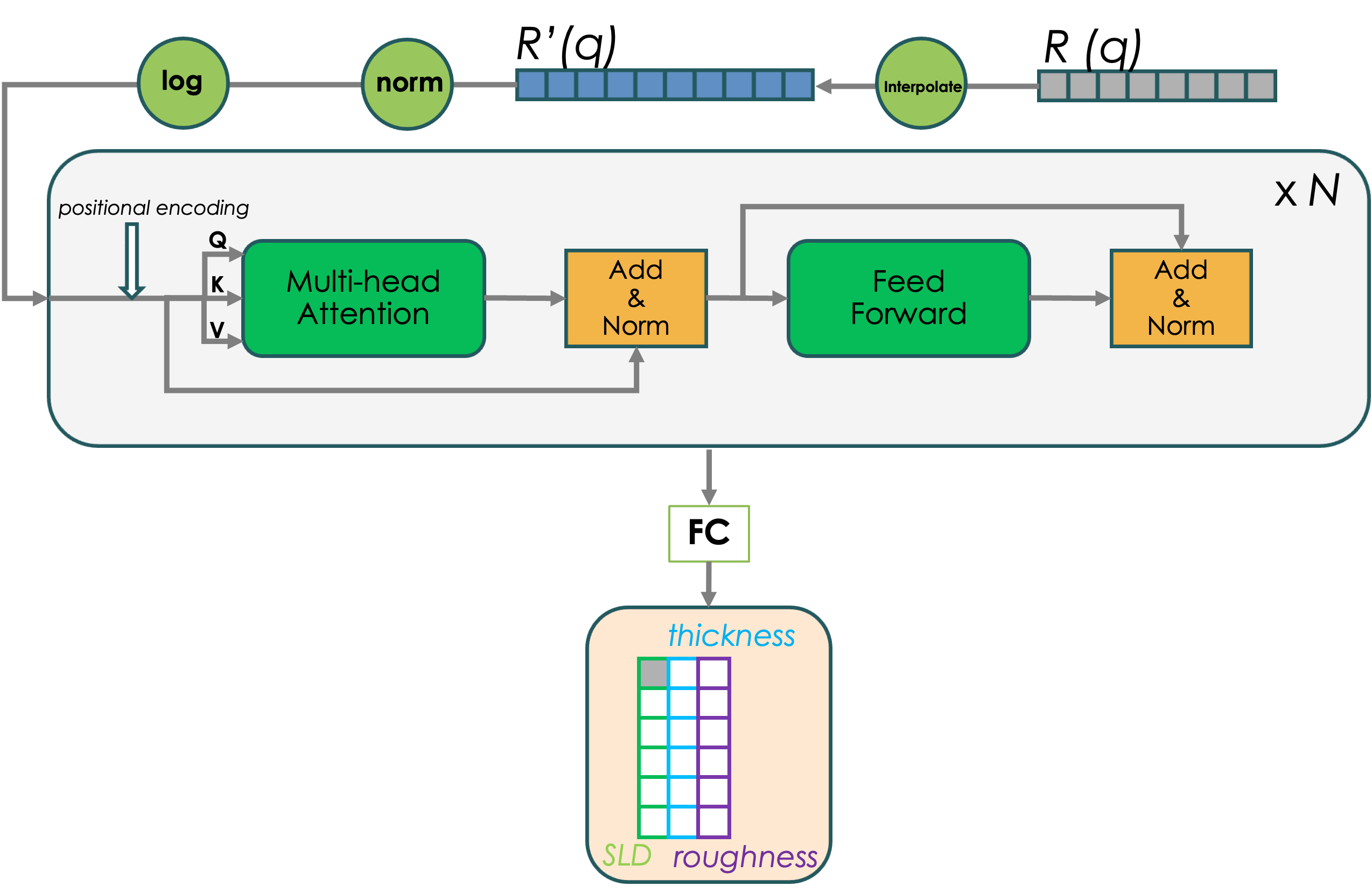}
\caption{%
\revision{Schematics of the \ModelName{} model, which consists of a pre-processing unit (top), transformer encoder based translation unit (middle) and a single fully connected layer (FC) for post-processing.}}
\label{fig:ntrace_schematic}
\end{figure}

\ModelName{} is constructed around a combined transformer encoder and \revision{a simple linear} \rrevision{encoder-decoder model as shown in the overall workflow chart \Cref{fig:ntrace_schematic}, which outlines the entire data processing pipeline—from raw measurements to refined structure prediction.}
\rrevision{\Cref{fig:schematic_multihead_attention} zooms in on the core component of this workflow, the transformer-based model, illustrating the multi-headed attention mechanism.
}
\revision{As illustrated in \Cref{fig:schematic_multihead_attention}, the pre-processed and position-encoded reflectivity profile is ingested by a transformer encoder.
Within this encoder, each attention head independently learns which specific portion of the reflectivity curve to focus on, using its learnable attention matrix to selectively extract relevant features while ignoring less pertinent information.
This process results in a feature vector that represents a unique aspect of the input data.
}

\begin{figure*}[htp]
\resizebox{1.8\columnwidth}{!}{
\begin{tikzpicture}[every node/.style={font=\sffamily}, node distance=2cm, >=stealth]

    \node (input) [draw, minimum width=3cm, minimum height=1cm, fill=blue!20] {Reflectivity Profile};

    \node (attention1) [right=2cm of input, draw, fill=green!20, rounded corners, minimum width=2cm, minimum height=1cm] {Attention Head 1};
    \node (attention2) [below=0.5cm of attention1, draw, fill=green!20, rounded corners, minimum width=2cm, minimum height=1cm] {Attention Head 2};
    \node (attention3) [below=0.5cm of attention2, draw, fill=green!20, rounded corners, minimum width=2cm, minimum height=1cm] {Attention Head 3};

    \node (feature1) [right=2cm of attention1, draw, fill=orange!20, minimum width=2cm, minimum height=1cm] {Feature 1};
    \node (feature2) [below=0.5cm of feature1, draw, fill=orange!20, minimum width=2cm, minimum height=1cm] {Feature 2};
    \node (feature3) [below=0.5cm of feature2, draw, fill=orange!20, minimum width=2cm, minimum height=1cm] {Feature 3};

    \node (pooling) [right=2cm of feature2, draw, fill=yellow!20, circle, minimum size=1.5cm] {Pooling};

    \node (decoder) [right=2cm of pooling, draw, fill=red!20, minimum width=2cm, minimum height=1cm] {Decoder};

    \node (output) [right=2cm of decoder, draw, fill=blue!20, minimum width=3cm, minimum height=1cm] {SLD Profile};

    \draw[->] (input) -- (attention1);
    \draw[->] (input.east) -- ++(0.75,0) |- (attention2);
    \draw[->] (input.east) -- ++(0.75,0) |- (attention3);
    \draw[->] (attention1) -- (feature1);
    \draw[->] (attention2) -- (feature2);
    \draw[->] (attention3) -- (feature3);
    \draw[->] (feature1.east) -- ++(0.75,0) |- (pooling.west);
    \draw[->] (feature2) -- (pooling);
    \draw[->] (feature3.east) -- ++(0.75,0) |- (pooling.west);
    \draw[->] (pooling) -- (decoder);
    \draw[->] (decoder) -- (output);

    \node[below=0.5cm of attention3] {\small Transformer Encoder};
    \node[below=0.5cm of feature3] {\small Extracted Features};
    \node[below=0.5cm of decoder] {\small Prediction of SLD Profile};

\end{tikzpicture}
}
\caption{%
\revision{
Schematic of the transformer-based model architecture for predicting the scattering length density (SLD) profile from reflectometry (NR) data.
}
}%
\label{fig:schematic_multihead_attention}
\end{figure*}

\revision{
Typically, a larger hidden dimension of the transformer encoder, $d_\text{model}$, and more layers, $n_\text{encoder layers}$, result in a more expressive model.
Similarly, a greater number of heads, $n_\text{head}$, allows the model to track more patterns simultaneously.
}
However, although larger models tend to be more expressive, it is not practical to use them for every application due to the increased computational cost.
It is better to find a balance between model size and correctness.
For example, a smaller model that fits easily on a single graphics card is often more desirable than an extremely large model that provides only a slight improvement in prediction accuracy.

To achieve this balance, hyperparameter tuning with Optuna~\cite{optuna_2019} was employed.
During the optimization process, a customized model size search range was provided to ensure that the model could fit on a single graphics card.
Optuna was then used to find models that provided the best results (lowest test loss with synthetic data).
After training 1000 cases, the model parameters with the smallest size were selected from the top 10 candidates, ensuring that the model could converge well without excessive size.
The final selected model parameters are listed in \Cref{table:params}.

\begin{table}[ht]
\centering
\caption{Training and model parameters}
\label{table:params}
\begin{tabular}{lr}
\toprule
\textbf{Parameter}            & \textbf{Value}                    \\
\midrule
\multicolumn{2}{c}{\textbf{Model Parameters}}                     \\
\midrule
$d_\text{model}$              & \num{2048}                        \\
$n_\text{head}$               & \num{32}                          \\
$n_\text{encoder layers}$     & \num{6}                           \\
$d_\text{input}$              & \num{150}                         \\
$d_\text{output}$             & \num{17}                          \\
\midrule
\multicolumn{2}{c}{\textbf{Training Parameters}}                  \\
\midrule
$n_\text{epochs}$             & \num{100}                         \\
$n_\text{data}$               & \num{1000000}                     \\
$n_\text{batch}$              & \num{150}                         \\
learning rate                & \num{5.793e-03}                   \\
weight decay                 & \num{7.199e-07}                   \\
optimizer                     & SGD                               \\
loss                          & hybrid                            \\
\bottomrule
\end{tabular}
\end{table}

\subsection{Loss Function and Optimization}

\ModelName{} uses a hybrid loss function that combines the discrepancies between extracted parameters and reference parameters, $L(r)_\text{structure}$, and the differences in corresponding reflectivity curves generated using these parameters, $L(r)_\text{scattering}$, i.e.
\begin{equation}
    L(r) = L(r)_\text{structure} + L(r)_\text{scattering}
\end{equation}
This dual objective design is used because machine learning models tend to be biased towards the parameter space covered during the training, which can result in limited generalization capability.
In this study, the additional loss term $L(r)_\text{scattering}$ ensures that the model can maintain accurate predictions when the truth lies outside the range of the training data.

It is also observed that most experimental measurements tend to have outliers due to various issues.
As a result, a L1 loss, specifically SmoothL1Loss from Pytorch, is used to calculate the differences in parameters and reflectivity curves, making the model's predictions resistant to large outliers.
The Stochastic Gradient Descent (SGD) optimizer in PyTorch, known for its stability, is used in this study.
The corresponding training parameters can be found in \Cref{table:params}.

\subsection{Training, validation, and evaluation}
\Cref{tab:data_range} shows the parameter ranges used when generating the synthetic training and validation data.
Performance evaluations of these models will demonstrate their expressiveness when applied to their intended systems and their ability to generalize to systems they were not trained on.
Throughout the training process, a curated set of 51 measurements were used to assess training.
The training process involves the following stages:

\begin{itemize}
    \item \textbf{Training with synthetic data:}
    In each epoch, the model is first trained on a synthetic dataset.
    The loss calculation involves two components: the structure loss, which is the difference between predicted parameters and reference parameters, and the scattering loss, which is the difference between the reflectivity curves generated using predicted parameters and those generated with reference parameters.
    These are combined to form the total loss, which is used to update the model weights.

    \item \textbf{Validation with synthetic data:}
    Following each training epoch, validation is performed using a separate synthetic validation dataset.
    The loss computation remains the same as in the training phase but without the need to track gradients, as this is essentially a testing step to monitor overfitting and model performance under controlled conditions.

    \item \textbf{Real data evaluation:}
    After the validation phase, the model is evaluated using the curated dataset derived from real experimental measurements.
    The input reflectivity curves are first interpolated to a standard $q$ range before being fed into the \ModelName{} model. 
    The model's predictions for sample parameters are then compared with reference parameters manually extracted by domain experts.
    The error of the prediction is quantified using the metric $\sqrt{\Delta p^2}$, which evaluates the mean squared difference between the predicted ($p_\text{pred}$) and reference parameters ($p_\text{ref}$).
\end{itemize}

The entire training, validation, and testing process was managed with mlflow to ensure all generated data were correctly labeled and documented for subsequent analysis.
The best-performing model, as determined by training loss, validation loss, and evaluation loss, is periodically saved.

\begin{table*}[ht]
\centering
\caption{Parameter ranges for the training data and the automated refinement.
        The ranges for the automated refined include expert knowledge of the system
        to assess valid ranges. The parameters $p_i$ represent the predicted parameter
        value for parameter $i$.}
\label{tab:data_range}
\begin{tabular}{llcc}
\toprule
\textbf{Number} & \textbf{Parameter}     & \textbf{Range} & \textbf{Refinement range}\\
\midrule
1  & Electrolyte SLD ($10^{-6}$\si{\per\angstrom\squared}) 
   & \SIR{5.0}{7.0}    & \SIR{6.0}{7.0} \\

2  & Electrolyte roughness  (\si{\angstrom})
   & \SIR{5}{120}      & 0.1$p_2$--1.5$p_4$ \\

3  & SEI SLD ($10^{-6}$\si{\per\angstrom\squared})
   & \SIR{-5.0}{6.5} & \SIR{3.5}{6.5} \\

4  & SEI thickness (\si{\angstrom})
   & \SIR{10}{500} & 100--2$p_4$ \\

5  & SEI roughness (\si{\angstrom})
   & \SIR{1}{80} & 0.1$p_4$--$p_7$/2\\

6  & Inner layer SLD ($10^{-6}$\si{\per\angstrom\squared})
   & \SIR{-2}{6} & 0.5$p_6$--1.5$p_6$\\

7  & Inner layer thickness (\si{\angstrom}) 
   & \SIR{10}{200} & 10--2$p_7$\\

8  & Inner layer roughness (\si{\angstrom}) 
   & \SIR{1}{35} & 0.1$p_8$--$p_7$/2.5\\

9  & Copper SLD ($10^{-6}$\si{\per\angstrom\squared})
   & \SIR{6}{7} & \SIR{6}{7} \\

10 & Copper thickness (\si{\angstrom}) 
   & \SIR{20}{700} & 0.5$p_{10}$--1.5$p_{10}$  \\

11 & Copper roughness (\si{\angstrom}) 
   & \SIR{1}{35} & 0.5$p_{11}$--1.5$p_{11}$  \\

12 & Titanium SLD ($10^{-6}$\si{\per\angstrom\squared})
   & \SIR{-3.5}{0} & 0.5$p_{12}$--1.5$p_{12}$\\

13 & Titanium thickness (\si{\angstrom})
& \SIR{10}{100} & 0.1$p_{14}$--2$p_{14}$\\

14 & Titanium roughness (\si{\angstrom})
   & \SIR{1}{35} & 0.1$p_{14}$--2$p_{14}$\\

15 & SiOx SLD ($10^{-6}$\si{\per\angstrom\squared})
   & \SIR{1.0}{4.2} & 0.5$p_{15}$--1.5$p_{15}$\\

16 & SiOx thickness (\si{\angstrom}) 
   & \SIR{5}{50} & 0.1$p_{16}$--50\\

17 & SiOx roughness (\si{\angstrom})& \SIR{1}{10} & \SIR{1}{8} \\

\bottomrule
\end{tabular}
\end{table*}

\section{Results}

This section details \ModelName{}'s performance, beginning with its learning dynamics on synthetic data, followed by its application to real experimental measurements.
The analysis covers initial predictions, refinement processes, and overfitting, concluding with an exploration of the model's zero-shot generalization capabilities.

\begin{figure}[ht]
\centering
\includegraphics[width=240px]{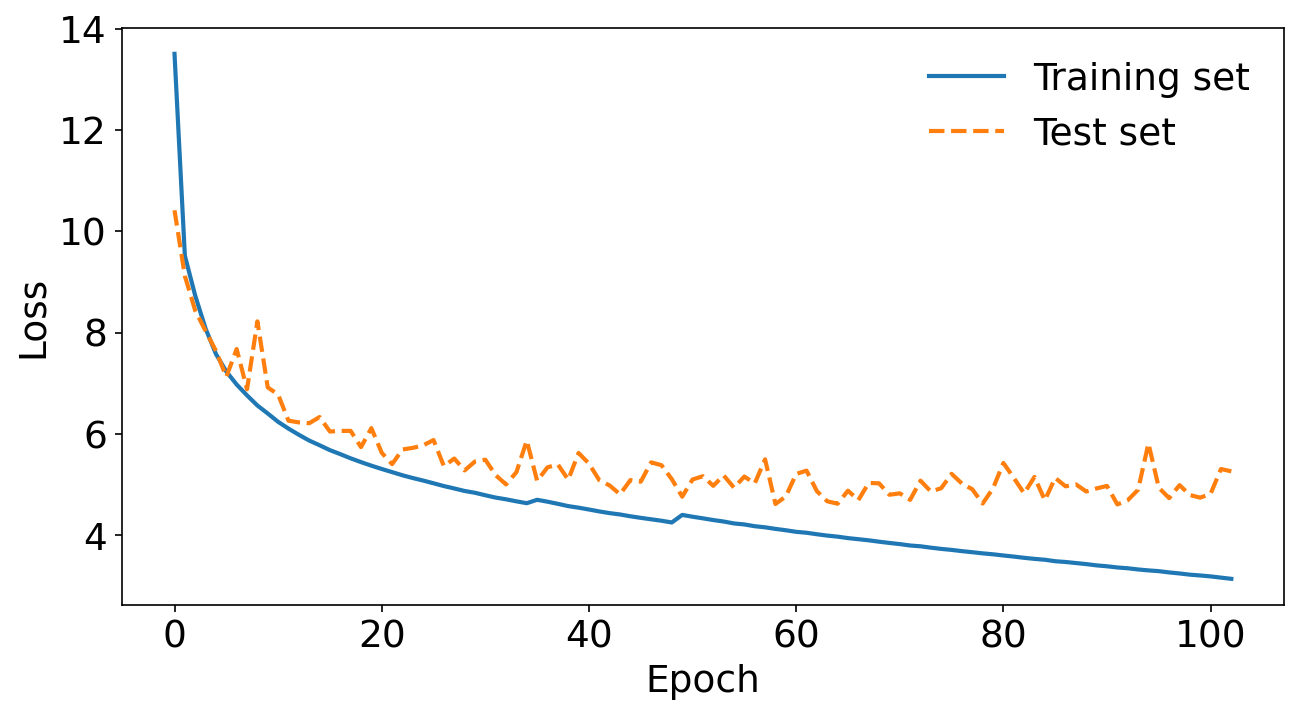}
  \caption{%
  Training and testing loss.
  }
  \label{fig:narrow-loss}
\end{figure}

The \ModelName{} model's learning dynamics, when trained on synthetic data, undergo distinct phases of adaptation and optimization, as shown in \Cref{fig:narrow-loss}.
Initially, both training and testing losses decrease rapidly, reflecting the model's swift adaptation to the solution space until around the 10th epoch.
After this point, until the 20th epoch, the rate of loss reduction significantly slows, suggesting a transition from rapid learning to more nuanced refinements as the model nears optimal solutions.
Beyond the 20th epoch, the training loss continues to decrease, approaching a value of one, while the testing loss stabilizes but becomes more variable.
This discrepancy signals the presence of potential over-fitting problems, as the model begins to adapt more to the training data's specifics rather than to generalize to new data.

The primary role of the \ModelName{} model is to facilitate efficient and accurate analysis of reflectometry measurements.
As such, it is vital to evaluate the model's performance against expert-labeled experimental measurements at different training stages.
To this end, the model was tested with measurements from structures akin to those used in its training.
Initially, sample parameters are obtained through direct inference by inputting the experiment's reflectivity measurements into \ModelName{}.
These direct inference results are then used as an initial guess for an automated refinement procedure using refl1d, aiming to enhance the accuracy and precision of the results further.
\Cref{tab:data_range} shows the refl1d fit ranges, which include domain knowledge of the system.
One of the main constraints is the SLD of the copper layer.
That constraint alone was found to be effective in excluding unphysical solutions.

\begin{figure}[ht]
\centering
\includegraphics[width=240px]{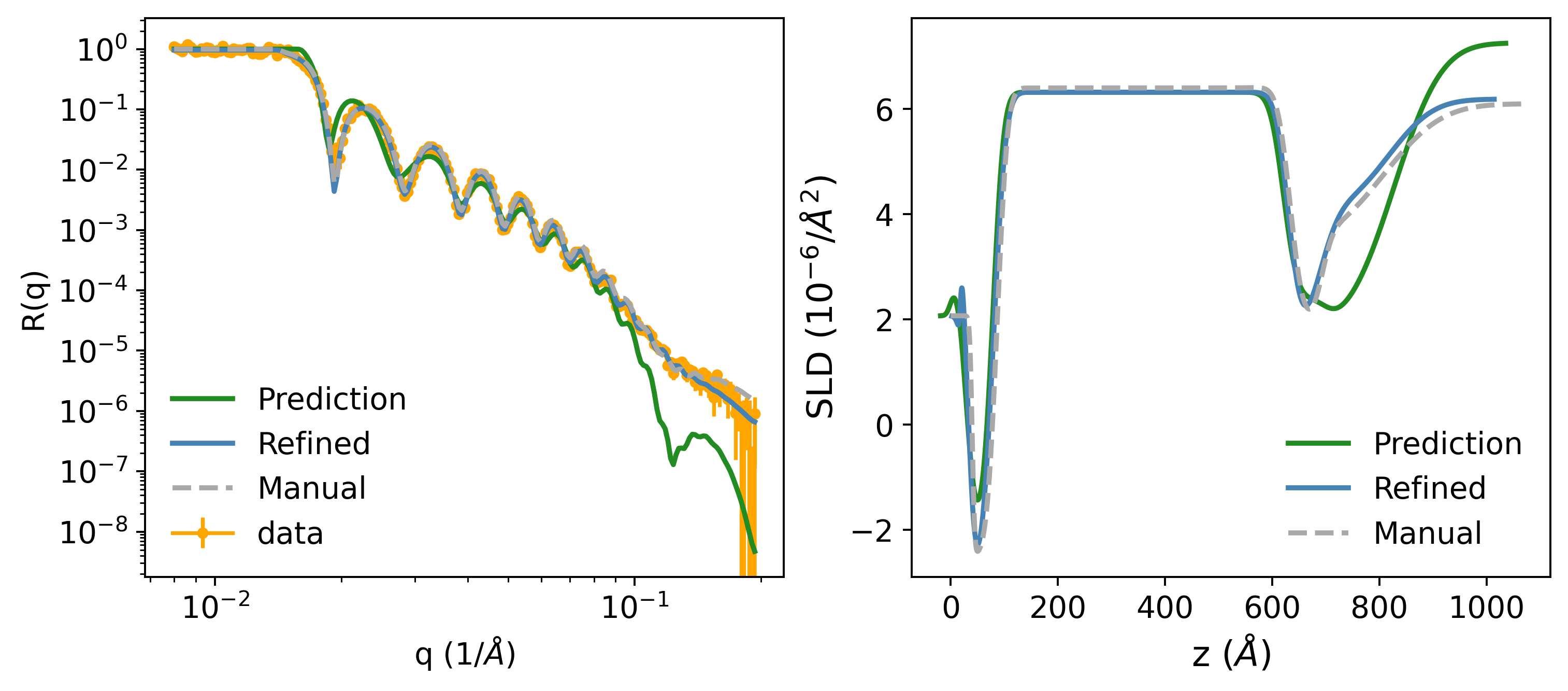}
  \caption{%
    Example reflectivity (left) and SLD profile (right) comparison between experiment measurements (data), direct inference from \ModelName{} (Prediction), augmented results via refinement (Refined), and independent expert-labeled results (Manual).
  }
  \label{fig:narrow-Cu-1}
\end{figure}

An example of this process is illustrated in \Cref{fig:narrow-Cu-1}, which displays the model's performance on a real experimental measurement.
The direct inference from \ModelName{} already aligns well with the raw data in the reflectivity curve and the corresponding SLD profile, indicating that \ModelName{} is capable of providing good initial guesses.
The additional automated refinement, where parameters were left to vary by up to a factor of two, mainly fine-tunes parameters to better match local features in both the reflectivity curves and the SLD profile.
Since the initial guess provided by \ModelName{} is already close to the true solution, the fitting process now takes much less time when compared with conventional approaches.
\Cref{fig:narrow-Cu-1} also shows the reference model obtained independently by a domain expert using refl1d.
\rrevision{%
The effectiveness of \ModelName{} in providing accurate initial guesses can be attributed to the multi-head attention mechanism's ability to focus on diverse aspects of the input reflectivity curves. 
A detailed explanation and a comparison of this mechanism with a CNN-based model are available in the Supplementary Information.
}

\begin{figure}[ht]
\centering
\includegraphics[width=240px]{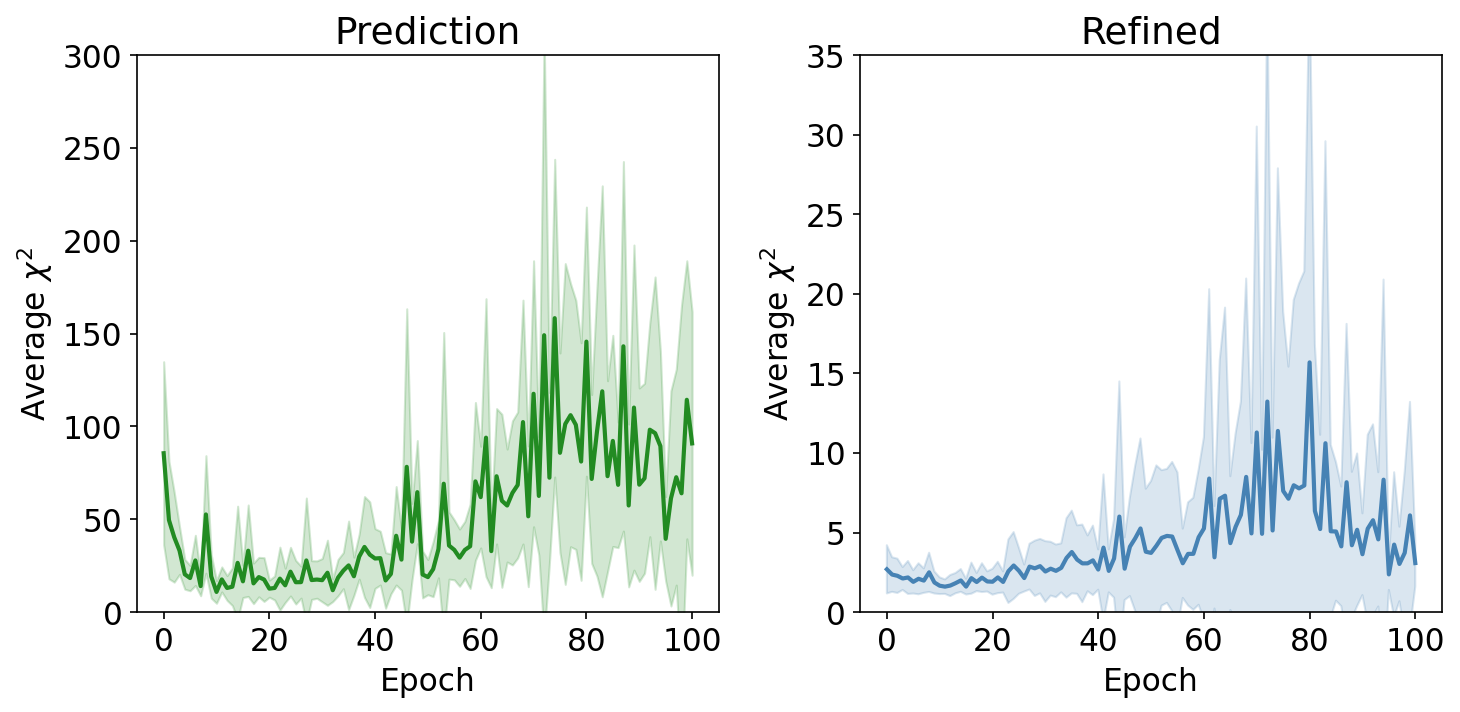}
  \caption{%
    Average $\chi^2$ of sample parameters from inference with \ModelName{} (left y-axis, green) and inference plus automated refinement (right y-axis, blue). The shaded area 
    represents the standard deviation of the $\chi^2$ distribution for all measurements.
    }
  \label{fig:narrow-Cu-chi2}
\end{figure}

These observations are also valid for other real data cases, which include 51 expert-labeled experiment measurements.
The performance of \ModelName{} is demonstrated through $\chi^2$ analysis, as shown in \Cref{fig:narrow-Cu-chi2}.
In general, the refinement process significantly improves the accuracy of the structure parameters, but its effectiveness depends on the quality of the initial predictions.
Over time, particularly after the 40th epoch, a noticeable decrease in the model’s predictive accuracy suggests over-fitting to the synthetic data.
This trend, initially undetectable in the synthetic data analysis, later becomes evident, emphasizing the need for careful interpretation of extended training on synthetic datasets.
We interpret this worsening of the predictions to the fact that synthetic data can never fully
capture the complexity of real measurements. Real measurements will contain instrumental
effects, some known and some unknown, which may lead to systematic shifts in the data. After
a large number of epochs, we reach a point where the model is learning detailed features in
the synthetic data that are no longer representative of real measurements. As we refine our
instrument simulation, we expect this effect to happen after a longer training period.

\begin{figure}[ht]
\centering
\includegraphics[width=240px]{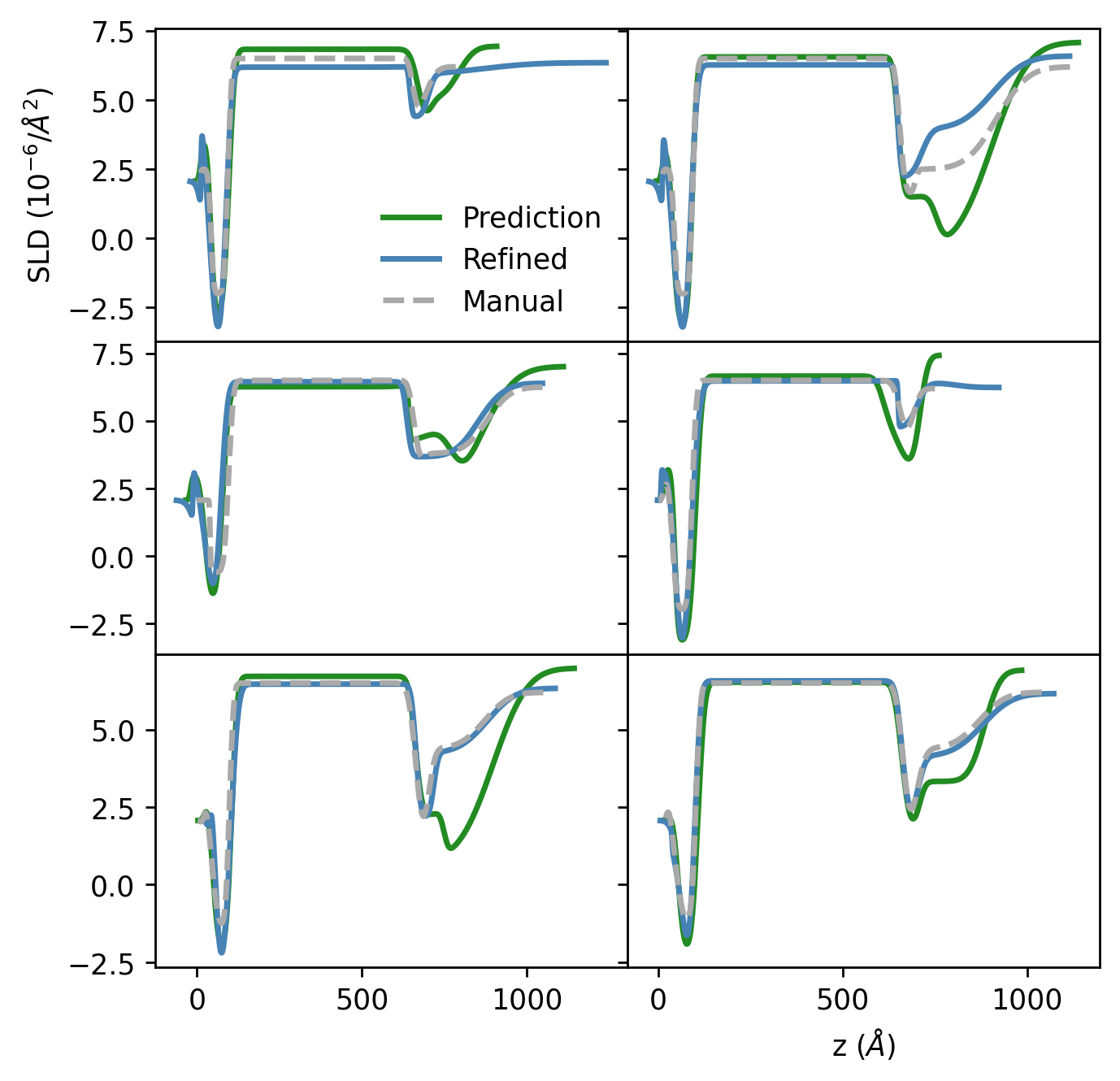}
  \caption{%
    Examples of SLD profiles for different measurements. The direct inference from \ModelName{} (Prediction), automated refinement (Refined), and independent expert-labeled results (Manual) are shown.
    \revision{
    All six subplots share the same x-axis and y-axis labels and units.
    The x-label is positioned in the bottom-right subplot, and the y-label is placed in the top-left subplot.
    A shared legend is located in the top-left subplot.
    }
  }
  \label{fig:six-examples}
\end{figure}

\Cref{fig:six-examples} shows the predictions and refined predictions for six distinct measurements. Each refined model was obtained automatically using the parameter ranges from
\Cref{tab:data_range}. The procedure allows the refinement to recover from unphysical
predictions. In all cases the prediction is close enough to produce a reasonable refined
model that is close to the manually obtained fit. Although certainly not perfect, these
results show the kind of predictions that can be expected using our approach. These 
predictions could be further improved with more precise knowledge of the samples measured.
In this analysis, each measurement was considered in isolation, whereas measurements on the
same sample will generally further constrain the allowable parameter ranges. 
The approach shown here would allow users to rapidly obtain a physical prediction that could
easily be adjusted to produce final results.

A zero-shot study was also carried out to examine the generalization capabilities of the \ModelName{} model, tested on entirely different systems without additional training.
The goal was to see if the model could adapt its learned parameters to new, unseen systems through an extended training dataset.

\begin{figure}[ht]
\centering
\includegraphics[width=240px]{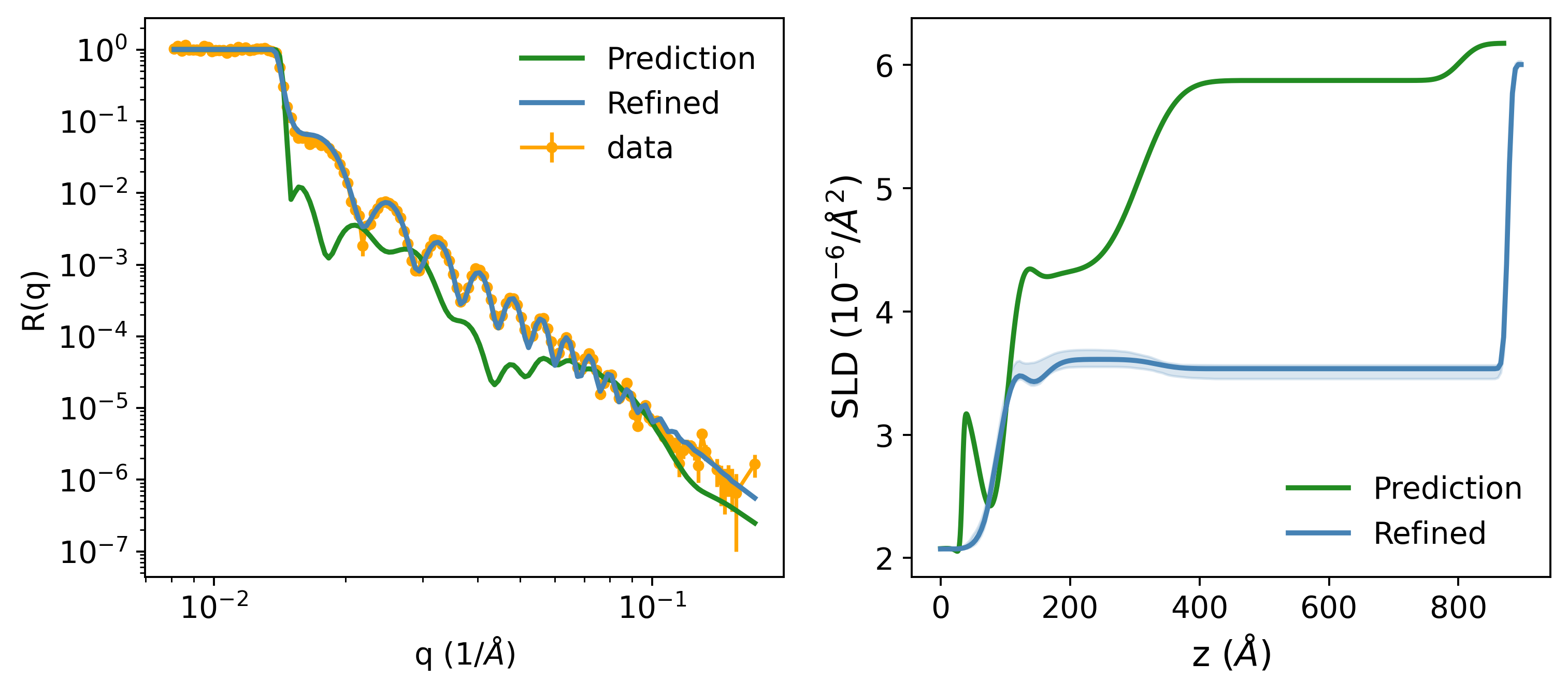}
\caption{%
    Example reflectivity (left) and SLD profile (right) comparison between experiment measurements of a single molybdenum film (data), direct inference from N-TRACE (Prediction), and augmented results via refinement (Refined). The shaded area
    represents the 90\% confidence level interval of the refl1d refinement.
    }
\label{fig:narrow-Mo-1}
\end{figure}

Figure~\ref{fig:narrow-Mo-1} illustrates the \ModelName{} prediction for a measurement of a single molybdenum layer film in the same electrochemical cell using a similar THF-based electrolyte.
The overall prediction was mostly inadequate.
Given the sharp critical edge and the SLD range of the synthetic data, finding a good SLD for the electrolyte is expected.
The refined prediction, for which the SLD ranges were allowed to vary over a wider range to ensure coverage of the SLD of molybdenum, surprisingly showed consistency with a single layer, featuring an SLD appropriate for molybdenum. 
This refinement also indicated that the total thickness of the predicted structure was similar to the refined thickness.

While the observation above indicates that an information-adding step in the automated refinement can in principle help recover from a poor initial prediction, we do not deem the quality of the prediction
to be good enough to be usable in production, especially in cases more complex than a single layer. \revision{Additionally, although \ModelName{} may be able to make predictions for simpler structures with fewer layers, it cannot be expected to produce reasonable results for structures more complex than those it was trained for. For instance, any structure needing to be modeled with more than two layers on top of the copper would not be a good use case for using our trained model.}
These findings suggest that although \ModelName{} can provide a basic level of generalization, it is not capable of discerning difference systems from input reflectometry measurements alone.
The \ModelName{} architecture requires incorporating specific data characteristics of the target system to achieve a prediction accuracy good enough for subsequent automated
refinement.
This underscores the need for a hybrid approach, combining general-purpose learning with system-specific adjustments to fully utilize the model's potential across various applications.
In other words, a mixture of experts (MOE)~\cite{masoudnia_moe_2014,chen_moe_2022} type model repository where each \ModelName{}-based expert trained for a specific system is probably the most feasible approach to utilize \ModelName{} in production.

\section{Summary}
This study highlights the potential of the \ModelName{} model, a neural network that utilizes the transformer architecture, for simplifying the analysis of neutron reflectometry data.
\ModelName{} notably reduces data analysis time by providing accurate initial guesses and enabling efficient automated refinements.
Rigorous testing on both synthetic and real experimental datasets has demonstrated the model's ability to adapt swiftly and effectively within the trained system's specific constraints.
However, a zero-shot generalization study revealed the model's limitations in performing across vastly different systems without retraining, indicating a need for further improvements in its generalization capabilities.

Moving forward, the development of \ModelName{} will focus on incorporating graph neural networks (GNN) based encoders as well as a hybrid architecture to extend its generalization capability.
On the encoder front, a GNN based encoder should allow the model to ingest raw data without any pre-processing, further simplify the data analysis process.
On the hybrid architectural front, MOE-type model repository will improve \ModelName{}'s generalization ability significantly, making it more adaptable and effective in various experimental setups and conditions.
By continually refining \ModelName{}, the objective is to democratize access to advanced analytical capabilities, thereby accelerating scientific discovery and innovation in materials science and beyond.

\begin{acknowledgement}
A portion of this research used resources at the Spallation Neutron Source (SNS), a Department of Energy (DOE) Office of Science User Facility operated by Oak Ridge National Laboratory.
Neutron reflectometry measurements were carried out on the Liquids Reflectometer at the SNS, which is sponsored by the Scientific User Facilities Division, Office of Basic Energy Sciences, DOE.
This research also used birthright cloud resources of the Compute and Data Environment for Science (CADES) at the Oak Ridge National Laboratory, which is supported by the Office of Science of the U.S. Department of Energy under Contract No. DE-AC05-00OR22725.
V.A.N., P.B., and T.F.J. acknowledge funding from the Villum Fonden (V-SUSTAIN grant 9455). V.A.N. was supported under the National Science Foundation Graduate Research Fellowship Program under grant no. DGE-1656518 and the Camille and Henry Dreyfus Foundation. V.A.N., P.B. and T.F.J. were supported by the U.S. Department of Energy, Office of Science, Office of Basic Energy Sciences, Chemical Sciences, Geosciences, and Biosciences Division, Catalysis Science Program through the SUNCAT Center for Interface Science and Catalysis.

\end{acknowledgement}

\rrevision{

\begin{suppinfo}
Explanation of the transformer's ability to capture short-range and long-range features, and comparison between \ModelName{} and a convolutional neural network (CNN).
\end{suppinfo}

}


\bibliography{n-trace}

\end{document}


\section{%
Explanation on Transformer Capability in Capturing Short and Long Range Features
}

The ability of transformers to capture both short-range and long-range dependencies is a key advantage in our task of extracting structural parameters from reflectometry data.
As each head in the multi-head attention mechanism independently focuses on different portions of the reflectivity curve, the model can learn multiple, diverse patterns in parallel.
Recent work has shown that different types of attention heads, such as Parallel, Radioactive, and X-type, can have varied impacts on model performance, emphasizing the importance of their diversity in interpretability and generalization tasks~\cite{guo2024interpretability}.

\begin{figure*}[htp]
    \centering
    \includegraphics[width=\textwidth]{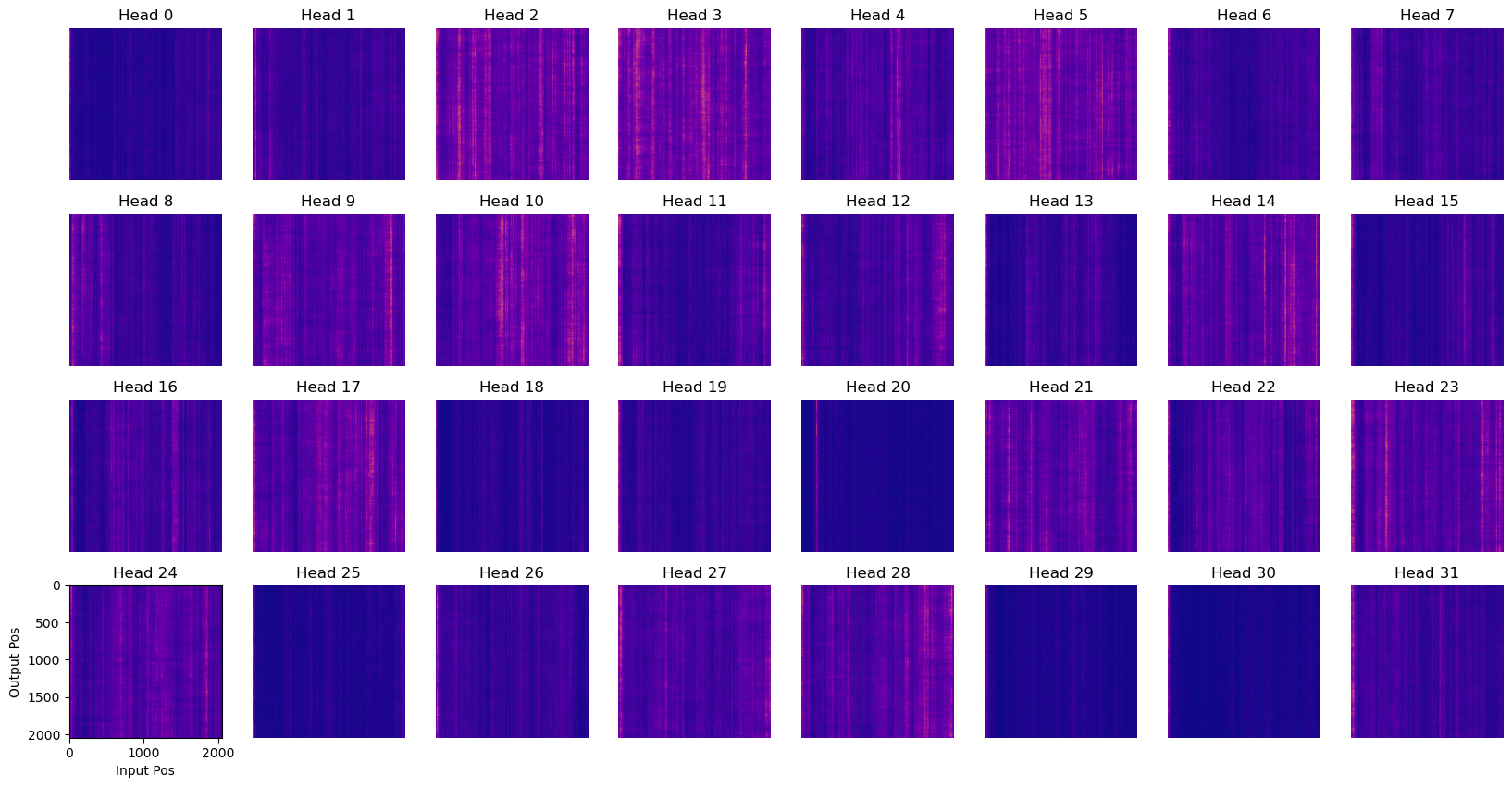}
    \caption{%
    Attention maps for the 32 heads in the first layer of the transformer model. Each subplot corresponds to a different attention head and visualizes how each head focuses on different sections of the input reflectivity sequence. The x-axis represents the input positions, while the y-axis represents the output positions in the latent space (sequence length of 2048). Heads like 0 focus on local patterns (short-range dependencies), whereas others like Head 11 capture global patterns (long-range dependencies). This diversity in attention patterns enables the model to learn both short- and long-range relationships simultaneously.
    }
    \label{fig:attention_map_layer_0}
\end{figure*}

For example, \Cref{fig:attention_map_layer_0} illustrates the attention maps for each head in the first layer of our 6-layer transformer model.%
\footnote{%
The attention map is dynamic and varies with different inputs, as it is calculated based on the specific input sequence.
The map shown here corresponds to one particular example input for demonstration purposes.
}
In this figure, we observe that different heads are responsible for learning unique dependencies.
Some heads, such as Head 0, focus on specific parts of the input sequence, capturing localized features (short-range relationships).
Meanwhile, others, like Head 11, exhibit a more global attention span, capturing dependencies between the beginning and end of the sequence (long-range relationships). 
Capturing these relationships would traditionally require deep Convolutional Neural Networks (CNNs), but the transformer architecture achieves this with significantly fewer layers.

Furthermore, as the reflectivity input is upscaled from 150 to 2048 dimensions using a linear encoder, the attention mechanism helps the model effectively map these latent representations to meaningful structural parameters.
The 32 attention heads, arranged in multiple layers, ensure a rich and comprehensive understanding of both local and global structural features, making the N-TRACE model suitable for this task.

\section{%
Comparison Study of a Convolutional Neural Network and NTRACE
}

To demonstrate how the transformer-based NTRACE model performs compared to traditional convolutional neural network (CNN) approaches, we conducted a comparative study using a conventional CNN model alongside fitting results performed using refl1d by a domain expert.
This section presents the methodology, results, and analysis of this comparison.

For this comparison, we implemented a CNN-based model designed to perform the same task as NTRACE: translating input reflectivity curves into layer parameters.
The CNN model consists of 1D convolutional layers for feature extraction, followed by fully connected layers for parameter prediction.
Here's a brief overview of its architecture:

\begin{itemize}
    \item
    Input layer: Accepts 1D reflectivity data (150 data points)

    \item
    Convolutional layers: Three 1D convolutional layers with increasing channel sizes (32, 64, 128), each followed by ReLU activation and max pooling

    \item
    Flatten layer: Converts the output of convolutional layers into a 1D vector

    \item
    Fully connected layers: Two dense layers (256 and 128 neurons) with ReLU activation

    \item
    Output layer: A final dense layer producing the 17 layer parameters

\end{itemize}

Both NTRACE and CNN models were trained on the same synthetic dataset and evaluated using the same set of real experimental measurements.

\begin{figure}
    \includegraphics[width=\linewidth]{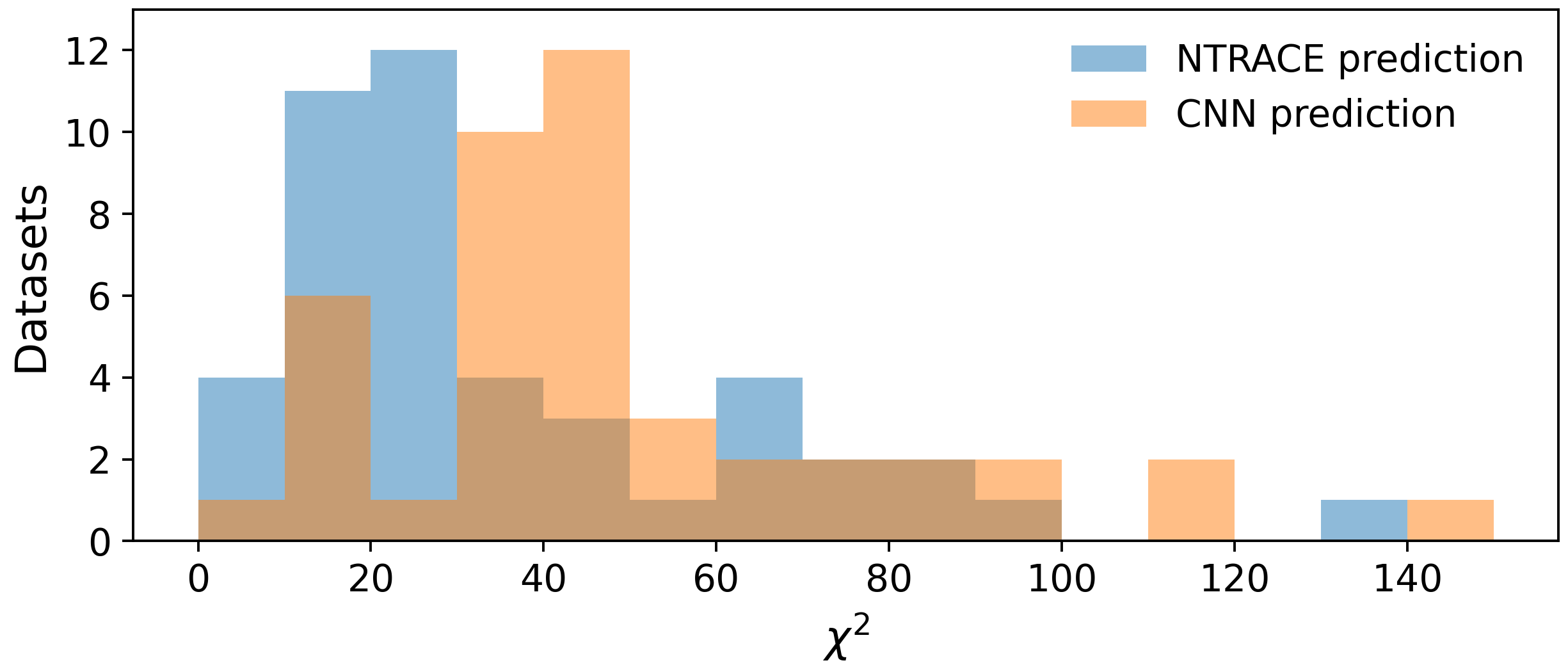}
    \caption{%
    $\chi^2$ distribution of NTRACE and CNN predictions for expert labeled data.
    }
    \label{fig:cnn_ntrace_manual_comparison}
\end{figure}

\Cref{fig:cnn_ntrace_manual_comparison} shows the $\chi^2$ distribution of the NTRACE and CNN predictions for the expert labeled real data, before final refinement. 
For most of the dataset, the NTRACE prediction fits the measurements better than our CNN.
It should be pointed out that one can always design a more complex CNN model to produce
better predictions. Such a comparison is only useful to point our that the 
transformer-based model performs well compared to models that are typically found in the
reflectometry literature, and that this performance is achieved with an expressive model
architecture that requires less optimization.

\bibliography{n-trace}

%% file: config.tex
\usepackage[export]{adjustbox}
\usepackage{siunitx}
\usepackage{amsmath}
\usepackage{xspace}
\usepackage{hyperref}
\usepackage{cleveref}
\usepackage{svg}
\usepackage{graphicx}
\usepackage{adjustbox}
\usepackage{booktabs} 
\usepackage{longtable} 
\usepackage{array} 

\newcommand{\SIR}[2]{\SIrange{#1}{#2}{}}

\newcommand{\QSIR}[3]{$#1 < #2 < #3$ \, \si{\per\angstrom}}
\newcommand{\ModelName}{\textit{N-TRACE}}
\DeclareSIUnit\angstrom{\text {Å}}

\usepackage{tikz}
\usetikzlibrary{shapes.geometric, arrows, positioning, fit, calc}
\tikzstyle{startstop} = [rectangle, rounded corners, minimum width=2cm, minimum height=0.5cm,text centered, draw=black, fill=red!30]
\tikzstyle{process} = [rectangle, minimum width=2cm, minimum height=0.5cm, text centered, text width=2cm, draw=black, fill=orange!30]
\tikzstyle{decision} = [diamond, minimum width=1cm, minimum height=0.25cm, text centered, draw=black, fill=green!30]
\tikzstyle{arrow} = [thick,->,>=stealth]